\documentclass[twocolumn,preprintnumbers,a4paper,superscriptaddress,
aps,showpacs,floatfix,nofootinbib]{revtex4}
\usepackage{graphicx}
\usepackage{bm}
\begin{document}
\def\la{\langle}
\def\ra{\rangle}
\newcommand{\beq}{\begin{equation}}
\newcommand{\eeq}{\end{equation}}
\newcommand{\beqa}{\begin{eqnarray}}
\newcommand{\eeqa}{\end{eqnarray}}
\newcommand{\intf}{\int_{-\infty}^\infty}
\newcommand{\into}{\int_0^\infty}

\title{Optical analog of 
Rabi oscillation suppression 
due to atomic   
motion}
\author{J. G. Muga}
\email[Email address: ]{jg.muga@ehu.es}
\affiliation{Departamento de Qu\'\i mica-F\'\i sica, Universidad del
Pa\'\i s Vasco, Apdo. 644, 48080 Bilbao, Spain}

\author{B. Navarro}
\email[Email address: ]{qfbnatob@lg.ehu.es}
\affiliation{Departamento de Qu\'\i mica-F\'\i sica, Universidad del
Pa\'\i s Vasco, Apdo. 644, 48080 Bilbao, Spain}


\begin{abstract}
The Rabi oscillations of a two-level atom illuminated by a laser on
resonance with the atomic transition 
may be suppressed by the atomic motion through averaging  
or filtering mechanisms. 
The optical analogs of these 
velocity effects are described.
The two atomic levels correspond in the optical analogy to orthogonal 
polarizations of light and the Rabi oscillations to polarization 
oscillations in a medium which is optically active, naturally or 
due to a magnetic 
field. In the later case, the two orthogonal polarizations could be
selected by choosing the orientation of the magnetic field,
and one of them be
filtered out.   
It is argued that the time-dependent optical polarization oscillations 
or their suppression are observable with current 
technology. 
\end{abstract}
\maketitle

\section{Introduction}
The analogies between phenomena occurring in two different physical systems  
open a route to find new effects  
or to translate solution techniques or devices, 
and quite often  
help to understand both systems better.
The parallelism between light and atom optics, in particular, 
has been a strong  
driving force for fundamental and applied research, 
and it has received recently 
renewed impulse with the advent of 
laser cooling techniques, and Bose-Einstein condensation of
alkali gases. Atomic interferometers or atom lasers are 
important examples that illustrate the fruitfulness 
of this correspondence.    

The analogy may
also enable us to perform in one system experiments 
which are difficult to carry out in the other one. 
As an example, 
tunneling time experiments are much easier for
microwaves than for 
matter waves \cite{tqm12,tqm11}.
Typically, optical analogs of quantum mechanical matter 
wave effects, if available, are simpler and 
less costly to implement than their matter counterparts. 
In addition, the electromagnetic pulses can be 
probed in a non-invasive way.

In this paper we shall describe the optical analogs of the 
time-dependent atomic Rabi oscillation in a laser field 
and of several dynamical suppression 
effects due to 
quantum, pure state atomic motion.
The Rabi oscillation is at the heart of 
measurement procedures for time and frequency standards \cite{Ramsey},
photon number in a cavity \cite{Haroche}, and other metrological
applications \cite{ai,DEHM02},
so its dynamic suppression may be a relevant effect, in  
particular for atomic clocks; 
This suppression has been also proposed
as a way to prepare  
specific internal atomic states by projection in quantum information 
applications 
\cite{NEMH03}.      

The relation between light and atom optics is frequently established 
at the level of translational (external) degrees of freedom, but here we need, 
in addition, an optical parallel of the two internal levels 
of the atom, which is provided by   
the orthogonal states of light polarization.   
Thus we follow in reverse order, from matter to light,  
a connection that can be traced back
historically 
to early experiments of I. Rabi,
later modified by N. F. Ramsey, which lead to the development 
of atomic clocks, 
and provided atomic analogs of polarization interferometry in optics,
with the internal states of the atom or molecule playing the role of
the polarization states of the photon
\cite{Ramsey}.

For completness a brief review of the dynamical suppression 
of the Rabi oscillation  
is provided in section II, which is mostly based on ref. \cite{NEMH03} 
but incorporates also some new elements; 
Section III is devoted to the description of the optical analog, 
Section IV presents numerical illustrations, and Section V describes
a possible implementation of polarization filtering 
making use of magnetic fields.

\section{Rabi oscillation suppression for moving atoms}

Neglecting decay, the effective Hamiltonian for a two level atom 
at rest and in presence of a detuned laser field 
is 
\beq
H= \frac{\hbar}{2} \left({0\atop \Omega^*}{\Omega
    \atop -2\delta} \right),
\label{ham}
\eeq
where $\Omega$ is the on-resonance Rabi frequency, 
$\delta=\omega_{L}-\omega$ is the detuning 
($\omega_{L}$ being the laser frequency and $\omega$
the atomic frequency)
and internal ground and excited states 
are represented as  
$|1\rangle \equiv {1 \choose 0}$ and $|2\rangle \equiv {0 \choose 1}$,
respectively. 
If at time zero the atom is
in the ground state, the ground and excited components 
of the state evolved with this Hamiltonian  are
\begin{eqnarray}
\psi^{(1)}&=&e^{i\delta t/2}[\cos(\Omega' t/2)-(i\delta/\Omega') 
\sin(\Omega't/2)],
\label{2}\\
\psi^{(2)}&=&e^{i\delta t/2}[(-i \Omega/ \Omega') \sin(\Omega' t/2)],
\label{3}
\end{eqnarray}
so that the populations $|\psi^{(1,2)}|^2$ 
alternate oscillating harmonically in time with 
``Rabi period'' 
$2\pi/\Omega'$ and effective Rabi frequency 
$\Omega'\equiv(\delta^2+|\Omega|^2)^{1/2}$.
The oscillation may however be suppressed when  
the atoms move into a region
illuminated by a perpendicular laser beam \cite{NEMH03}.  
For an idealized sharp laser profile in a one dimensional 
approximation (its validity and the three dimensional case 
are examined in \cite{HHM05}), the Hamiltonian becomes 
\beq
H= {\widehat{p}}^2/2m + \frac{\hbar}{2} \Theta({\widehat{z}})
\left({0\atop \Omega^*}
{\Omega
\atop -2\delta} \right), 
\label{H}
\eeq
where $\widehat{z}$ and $\widehat{p}$ are position and momentum operators. 
If reflection is negligible, for moderate to high velocities,
the stationary 
wave for incidence in the ground state with wavenumber
$k$ can be approximated, up to a normalization constant, 
as 
%

$\phi_k^{(1)}\approx e^{ikz} e^{i\delta t/2}\bigg[
\cos\left(\frac{\Omega' z m}{2 \hbar k}\right)-
(i\delta/\Omega')\sin\left(\frac{\Omega'z m}{2 \hbar k}\right)
\bigg],$

$\phi_k^{(2)}\approx e^{ikz}e^{i\delta t/2}\bigg[
(-i\delta/\Omega')\sin\left(\frac{\Omega' z m}{2 \hbar k}\right)
\bigg].$ 
\\
%
This may be interpreted semiclassically as the state of 
an ensemble of independent atoms which 
travel with momentum $k\hbar$ and sustain 
Rabi oscillations. Note that   
the quantity  $zm/\hbar k$ plays the 
role of time in the arguments of the trigonometric functions, 
compare with Eqs. (\ref{2}) and (\ref{3}). 
In other words, the Rabi oscillation is also 
evident spatially in the stationary waves, 
under the form of density undulations of the two
components  
with a 
``Rabi wavelength'' $\lambda_R=kh/m\Omega'$.

The Rabi oscillation may be suppressed ``adiabatically''
if 
a quasi-monochromatic, pure-state wave packet 
enters slowly 
into the laser illuminated region.  
Intuitively, and according to a classical 
picture, the atoms in the 
ensemble start to 
oscillate at different times because of their different entrance 
instants,  
so the global oscillation averages out 
if the entrance interval of the packet is greater than the Rabi
oscillation period.
This intuitive interpretation cannot be taken too literally though.
In particular, 
the suppression 
involves a pure state and not
a 
statistical mixture. Note also that no incoherent ``fading''
due to decay 
from the excited state is involved. (The effect of fading was 
discussed in \cite{NEMH03}.)

A second group of suppression effects is associated with state filtering 
or velocity splitting.
To explain these two related concepts, 
we need a more accurate representation than before.  
In the laser region,  
let $|\lambda_+\rangle$ and 
$|\lambda_-\rangle$ be the eigenstates,
corresponding to the eigenvalues $\lambda_\pm$, of the
matrix $\frac{1}{2} \left( {0\atop \Omega^*}{\Omega\atop -2\delta}
\right)$.  
One easily finds 
\begin{eqnarray}\label{A13}
\lambda_\pm &=&  \frac{-\delta \pm \Omega'}{2},
\\ \label{A14}
|\lambda_\pm \rangle &=&  {1 \choose \frac{-\delta \pm \Omega'}{\Omega}},  
\end{eqnarray} 
where $|\lambda_\pm \rangle$ have not been normalized.   
[For later comparison with ``circularly polarized states'' 
notice that by setting
$\delta=0$ and for the case $\Omega$=i$|\Omega|$,
$\lambda_\pm = \pm \frac{\Omega}{2}$,
and $|\lambda_\pm \rangle =  {1 \choose \mp i}$.] 
%
The stationary state for $z>0$ can be written as a superposition
\beq
\Phi_k(z)=C_+|\lambda_+\ra e^{ik_+z}+C_-|\lambda_-\ra
e^{ik_-z},
\eeq
where 
\beq
k_{\pm}=k\left(1 - \frac{m(-\delta \pm \Omega')}{\hbar k^2}\right)^{1/2},
\eeq
and the coefficients $C_\pm$ are obtained from the
matching conditions at $z=0$ \cite{NEMH03}. 
The two components $(\pm)$ have two different propagation velocities
and in fact the Rabi oscillation may be understood as 
an interference between the two terms when $k_\pm\approx
k + m(\delta \mp \Omega')/(2\hbar k)$. 
For initially quasimonochromatic packets, with  
the wavenumber spread much smaller than the average 
wavenumber, 
$\sigma_k/k_0\ll 1$, the two different velocities 
also imply eventually a spatial separation of the $|\lambda_\pm\ra$ 
components, so that the interference and associated Rabi oscilation 
finally disappear  
for times larger than the time to split the packet into two,   
$2\sigma_z k/\Omega'$, where $\sigma_z^2$ is
the spatial variance of the wave packet. 
An extreme case is the complete state filtering that occurs 
for very low kinetic energies, when  
$k_-$ is 
purely imaginary so that $e^{ik_-z}$ becomes an evanescent wave and 
only the $|\lambda_+\ra$ component survives 
in the laser region for $z$ greater than the penetration length
$[{\rm Im}(k_-)]^{-1}$.  
Since the exact form of the surviving $|\lambda_+\ra$ state 
can be modified by the laser detuning and Rabi frequency,
this state filtering effect provides a projection 
mechanism to prepare especific internal states 
regardless of the incident atomic state \cite{NEMH03}.

\section{Optical analog}
In the optical analog of the dynamic effects described, the internal
states will be substituted by orthogonal polarizations of the field,
and the laser 
region by an optically active medium; the rotation of the 
polarization plane  and corresponding oscillation of the 
linear polarization intensities will mimic the Rabi oscillation, 
and electromagnetic pulses will play the role of the atomic wave
packets.\footnote{Notice also that one further correspondence 
may be established with a spin-polarized electron 
incident on a region with a perpendicular magnetic 
field, 
i.e. there is an analogy between Rabi oscillations 
and Larmor precession, which has been used to extend
the concept of a Larmor clock \cite{tqm9,tqm8}
and other definitions of a traversal time to 
atomic systems \cite{70,72,73,tqm8,Gasparian,JK,DG,LLH}.} 
 
The use of a complex electric field 
for quasi-monochromatic 
pulses within analytic signal theory  facilitates 
the comparison and correspondence 
between quantum wave function and electric field components.    
Some of the parallelisms are quite direct. For example, 
we shall keep the same
notation for the quantum wavenumbers and the optical propagation constants 
$k$ and $k_\pm$, or for the coefficients in the stationary waves, 
such as $C_\pm$ and the reflection amplitudes $R$,  
although their values and detailed expressions need not be equal. 
Atomic populations in the ground state may be related to 
total energies in the vertical, linear polarization component and, 
similarly, the atomic excited state will be mimicked by horizontal, 
linear polarization. 
In our analogy, positions and times are present both in 
optics and quantum mechanics since we want to describe and compare 
pulses and wavepackets in space-time. 
This is at variance with the analogy 
established by Zaspasskii and
Kozlov \cite{ZK}, in which the spatial coordinate played the role 
of time in a Schr\"odinger-like  equation satisfied 
by the polarization vector.  
\subsection{Basic relations}
From the Maxwell equations in a nonmagnetic, 
electrically neutral dielectric medium, the wave equation for the 
${\bf E}$ field is 
\beq
\label{td}
\nabla\times(\nabla\times{\bf E})+\frac{1}{c^2}
\frac{\partial^2{\bf E}}{\partial t^2}
=-\mu_0\frac{\partial^2{\bf P}}{\partial t^2},
\eeq
where 
$\mu_0$ is the permeability of the vacuum, 
$c$ the speed of light in vacuum and 
${\bf P}$ the macroscopic polarization (volume density 
of electric dipoles), which we assume to be given by 
the linear constitutive relation 
\beq\label{p}
{\bf P}=\epsilon_0 {\bm \chi}{\bf E},
\eeq
where $\epsilon_0$ is the permittivity of the vacuum, 
and ${\bm{\chi}}$ is the susceptibility tensor
%
\beq
{\bm{\chi}}=\left[
\begin{array}{ccc}
\chi_{11}&\chi_{12}&\chi_{13}
\\
\chi_{12}^*&\chi_{22}&\chi_{23}
\\
\chi_{13}^*&\chi_{23}^*&\chi_{33}
\end{array}
\right].
\label{sus}
\eeq
%
The medium is assumed to be non-absorbing so that 
${\bm{\chi}}$ is hermitian. The absorbing case is briefly considered 
in section \ref{mo} and in the Appendix.

In the medium, $\bf{E}$ may not be perpendicular to the 
propagation direction but the electric displacement 
\beq\label{d}
{\bf D}=\epsilon_0 {\bf E}+{\bf P}
\eeq
is perpendicular, since, from Eq. (\ref{td}) and assuming a harmonic, 
plane wave solution $e^{i{\bf{k}}_m\cdot{\bf{r}}}e^{-i\omega t}$,   
\beq
\label{stati2}
{\bf k}_m\times({\bf k}_m\times{\bf E})
=-\frac{\omega^2}{c^2 \epsilon_0}{\bf D}. 
\eeq
The subscript $m$ in $k_m$ stands for ``medium'' and it will be used 
to avoid confusion with $k=\omega/c$ in vacuum. 
In components, and for ${\bf k}_m$ in $z$ direction,  $D_z=0$ and 
\beq
{ D_x}=\frac{c^2 \epsilon_0 k_m^2}{\omega^2}{ E_x},\;\;\;    
{ D_y}=\frac{c^2 \epsilon_0 k_m^2}{\omega^2}{ E_y}
\eeq
i.e., the displacement vector components whose ratio
determines the polarization state of the harmonic wave, 
are proportional 
to the corresponding components of {\bf E}.
From Eqs. (\ref{p}), (\ref{d}), and (\ref{stati2}),  
\beq
\label{stati}
{\bf k}_m\times({\bf k}_m\times{\bf E})+\frac{\omega^2}{c^2}{\bf E}
=-\frac{\omega^2}{c^2}{\bm\chi}{\bf E}, 
\eeq
which, written in components leads to
\beq
\label{ezcomp}
{E_{z}}=-(\chi_{13}^* {E_{x}}+\chi_{23}^* {E_{y}})/\gamma_3
\eeq
and the system
\beqa
\label{sistem1}
\!\!\bigg[\!\!-\!\!k_m^2\!+\!\frac{\!w^2\!}{\!c^2\!}\!\left(\gamma_1\!-\!
\frac{\!|\!\chi_{13}\!|^2\!}{\gamma_3}
\!\right)\!\!\bigg]\!E_{x}
\!\!+\!\!\frac{\!w^2\!}{\!c^2\!}\!\left(\!\chi_{12}\!-\!
\frac{\!\chi_{13}\chi_{23}^*\!}
{\gamma_3}\!\right)\!\!
E_{y}\!=\!0,\!
\nonumber\\
\!\!\frac{\!w^2\!}{\!c^2\!}\!\!\left(\!\chi_{12}^*\!-\!
\frac{\!\chi_{23}\chi_{13}^*\!}
{\gamma_3}\!\right)\!\!E_{x}
\!\!+\!\!\bigg[\!\!-\!k_m^2\!\!+\!\frac{\!w^2\!}{\!c^2\!}\!\left(\gamma_2\!-
\frac{\!|\!\chi_{23}\!|^2\!}
{\gamma_3}\!\right)\!\!\bigg]\!
E_{y}\!\!=\!0,
\eeqa
where $\gamma_j\equiv 1+\chi_{jj}$. 
It can be solved 
by making the determinant of the coefficients vanish. 
The result is a fourth order equation ($ k_m^4+\alpha k_m^2+ \beta $=0) 
with coefficients
%
\beqa
\alpha&=&\frac{-\omega^2}{c^2}\bigg[\gamma_1+\gamma_2
-\frac{|\chi_{13}|^2+|\chi_{23}|^2}
{\gamma_3}\bigg]
%
\\
\beta&=&\frac{\omega^4}{c^4}
\bigg[\gamma_1\gamma_2
-\frac{\gamma_2|\chi_{13}|^2}{\gamma_3}
-
\frac{\gamma_1|\chi_{23}|^2}{\gamma_3}
\\
&+&\frac{|\chi_{13}|^2|\chi_{23}|^2}{\gamma_3}
-\left|\chi_{12}-
\frac{\chi_{13}\chi_{23}}{\gamma_3}\right|^2\bigg]. 
\nonumber
\eeqa
Since we shall assume that the wave incides from a vacuum 
region adjacent to a semiinfinite medium, 
we only pick up two physical solutions (if complex they   
must have a positive imaginary part to decay; if real they must be positive
to implement outgoing boundary conditions)  
and denote them as $k_\pm$, 
\beq
k_\pm=\frac{\omega}{c}n_\pm, 
\eeq
$n_\pm$ being the corresponding refraction index. Examples will 
be given soon. 

The polarization corresponding to each solution 
in the $x-y$ plane, up to an intensity constant, 
may be given in terms of the Jones vector 
$\left[\begin{array}{c}E_{x}\\E_{y}\end{array}\right]_\pm$ 
\cite{Fowles}. 
For a closer comparison with the atomic case we may use  
the notation $E^{(1)}\equiv E_x$, $E^{(2)}\equiv E_y$, reminiscent 
of the atomic ground and excited state amplitudes.
In particular, 
the internal atomic states $({1\atop 0})$ (ground) and $({0\atop 1})$
(excited) correspond to the  
Jones vectors $\left[\begin{array}{c}1\\0\end{array}\right]$,
$\left[\begin{array}{c}0\\1\end{array}\right]$ which denote, respectively, 
``vertical'' ($x$-direction) and ``horizontal'' ($y$-direction) 
linear polarization. 

Since we are not specifying the total intensity, the 
polarization is also represented by any proportional vector. 
In particular, it is 
convenient to work with 
\begin{eqnarray}\label{A133op}
|\lambda_\pm \rangle &=&\left[\begin{array}{c}1\\(\frac{E_{y}}{E_{x}})_{\pm}
\end{array}\right]
\end{eqnarray} 
where
\beq
\label{exey}
(E_{y}/E_{x})_{\pm}=\frac{\gamma_1-\frac{|\chi_{13}|^2}{\gamma_3}-
\frac{k_{\pm}^2 c^2}{\omega^2}}{-\chi_{12}+{\chi_{13}\chi_{23}^*}/\gamma_3
}.
\eeq
Depending on the value of $(\frac{E_{y}}{E_{x}})_{\pm}=a+ib$ ($a$ and 
$b$ real), the
polarization can be linear ($b=0$), circular ($a=0$, $b=\pm 1$),
or in general elliptic.

For an optically active medium with a susceptibility tensor of 
the form 
\beq
{\bm{\chi}}=\left[
\begin{array}{ccc}
\chi_{11}&\chi_{12}&0
\\
\chi_{12}^*&\chi_{11}&0
\\
0&0&\chi_{33}
\end{array}
\right],\;\;\;{\rm{Re}}(\chi_{12})=0, 
\label{susfaraday}
\eeq
the optical propagation constants are given by 
\beq
\label{kmp}
k_\pm=\frac{\omega}{c}(1+\chi_{11}\pm|\chi_{12}|)^{1/2}.
\eeq
Assuming $\rm{Im}(\chi_{12})>0$,  
two orthogonal harmonic solutions of Eq. (\ref{td}), supplemented
by Eqs. (\ref{p}) and (\ref{sus}), for right ($+$) and 
left ($-$) circularly polarized light, forward-moving  or possibly 
evanescent (if $k_-$ becomes imaginary) are 
\beq
\left[\begin{array}{c}1\\-i\end{array}\right] e^{ik_+z} e^{-i\omega t},\;\;\;
\left[\begin{array}{c}1\\i\end{array}\right] e^{ik_-z} e^{-i\omega t},
\eeq
In vacuum, we have instead 
${\bm{\chi}}={\bf 0}$, and the following forward/backward moving 
orthogonal harmonic solutions (with signs $+/-$ respectively)
\beq
\left[\begin{array}{c}1\\0\end{array}\right] e^{\pm ikz} e^{-i\omega t},
\;\;\;
\left[\begin{array}{c}0\\1\end{array}\right] e^{\pm ikz} e^{-i\omega t},
\eeq
where $k=\omega/c>0$.

If the optically active medium occupies the region $z>0$,  
and for normal incidence (${\bf k}$ in $z$ direction)  
of vertically polarized light, 
the harmonic solution reads, up to a constant,
$e^{-i\omega t}{\bf{F}}_k(z)$, where 
\beqa
\label{medium}
{\bf{F}}_k(z)=
\!\!\left\{\begin{array}{l}
\left[\begin{array}{c}1\\0\end{array}\right]\!(e^{ikz}\!+\!R_{11}e^{-ikz}\!)
\!+\!
\left[\begin{array}{c}0\\1\end{array}\right]\!R_{21}e^{-ikz}, z<0 
\\
C_+\,\left[\!\begin{array}{c}1\\-i\end{array}\!\right] e^{ik_+z}
+
C_-\left[\begin{array}{c}1\\i\end{array}\right] e^{ik_-z}, z>0
\end{array}\right.
\eeqa
Imposing at $z=0$ the continuity of the tangential components 
of the electric and magnetic fields amounts, for normal incidence, 
to enforce the continuity 
of the components $E_{x,y}$ and their derivatives,   
\beqa
1+R_{11}&=&C_+ +C_-,
\\
k(1-R_{11})&=&k_+C_+ +k_-C_-,
\\
R_{21}&=&-iC_+ +iC_-,
\\
-kR_{21}&=&-ik_+C_+ +ik_- C_-.
\eeqa
Solving the system, 
\beqa
R_{11}&=&\frac{k_+k_--k^2}{(k+k_+)(k+k_-)},\;\:\:C_+=\frac{k}{k+k_+},
\\
R_{21}&=&\frac{-ik(k_--k_+)}{(k+k_+)(k+k_-)},\:\:\:C_-=\frac{k}{k+k_-}.
\eeqa
There are thus two propagation constants $k_\pm$, and two group velocities
in the optically active medium, 
\beq
\frac{d\omega}{dk_\pm}=\frac{c}{n_\pm}.
\eeq
When $|\chi_{11}|,\, |\chi_{12}|$ $<<1$,
the two propagation constants differ 
only slightly, 
\beq
k_\pm\approx \frac{\omega}{c}\left(1+\frac{\chi_{11}\pm|\chi_{12}|}{2}\right), 
\eeq
there is also negligible reflection, 
$R_{j1}\approx 0$ ($j=1,2$),
and $C_\pm\approx 1/2$, so that 
in the optically active medium, $z>0$, 
\beqa
\label{medium1}
E^{(1)}&\approx& e^{i(kz-\omega t)}\cos(k|\chi_{12}|z/2),
\\
\label{medium2}
E^{(2)}&\approx& e^{i(kz-\omega t)}\sin(k|\chi_{12}|z/2).
\eeqa
Each $z$ is thus characterized by some linear polarization 
that rotates when $z$ increases (optical activity). The  
rotation length for a full cycle of the moduli squared is  
$L=2\pi c/\omega|\chi_{12}|$ or, equivalently, the cycle requires a time 
$T=2\pi/\omega|\chi_{12}|$ for a quasimonochromatic pulse. 
Pulses are analogous to wave packets in the present
correspondence and are formed by superposition 
of harmonic components. 
Moreover, in the quasi-monochromatic regime, we can 
interpret the modulus squared of complex field components as 
(half) short time averages of the real field, according to 
the theory of complex analytic signals \cite{Mandel}.
Up to a constant that fixes the actual intensity,
the time dependent field is given by 
\beq\label{ext}
{\bf{E}}(z,t)=\frac{1}{(2\pi)^{1/2}}\int_0^\infty dk\, A(k)
{\bf{F}}_k(z) e^{-i\omega t}.  
\eeq
The time $\delta_t$ required, from
the entrance instant of the pulse peak, to  
split the pulse into  
two separated pulses of orthogonal circular polarization 
may be estimated by imposing that the 
spatial increment between the two components, due to their 
different group velocities, be equal to the pulse
width $\sigma_z$. In particular, for $\chi_{11}=0$ and $|\chi_{12}|<<1$,   
\beq
\label{st}
\delta_t=\frac{2 \sigma_t}{|\chi_{12}|},
\eeq
where $\sigma_t=\sigma_z/c$. 
A time dependent observation of the polarization rotation, 
which is the optical analog of a temporal Rabi oscillation, requires 
$\delta_t>T$ to avoid the splitting,
but also $\sigma_t<T$ to avoid an averaging suppression.  
Combining the two constraints gives the ideal conditions 
\beq
\chi_{12}<<\frac{\sigma_k}{k_0}<<1.
\eeq
%

%
%
%
%
%
%
%
%
%
%
%
\section{Numerical example}
In this section we shall demonstrate with numerical examples 
the time dependence of the 
polarization rotation in the optically active medium 
and several dynamical 
suppression effects.  
The pulse is chosen  
within the quasi-monochromatic regime, so  
that we may    
neglect any variation with $k$ of the matrix elements 
of the susceptibility and consider constant values in each 
calculation.   
We have used Eq. (\ref{ext})
with the Gaussian amplitude 
\beq\label{ak}
A(k)=\sqrt{\sigma_z}\left(\frac{2}{\pi}\right)^{1/4}
e^{-(k-k_0)^2\sigma_z^2}e^{-ikz_0}.
\eeq
It is for convenience ``normalized'' so that 
$\int dk\,|A(k)|^2=1$ and, at time zero,
$\int dz\, |E^{(1)}(z,t=0)|^2=1$. 
In all cases the central wavelength is chosen in the
visible region of the spectrum, 
$\lambda_0=2\pi/k_0=500$ nm. 
The pulse is thus a right moving Gaussian pulse of vertically
polarized light centered at time 
$t=0$ at $z_0<0$, outside the optically active medium, 
and with spatial variance $\sigma_z^2$.     
 
\begin{figure}[h]
{\includegraphics[width=3.3in]{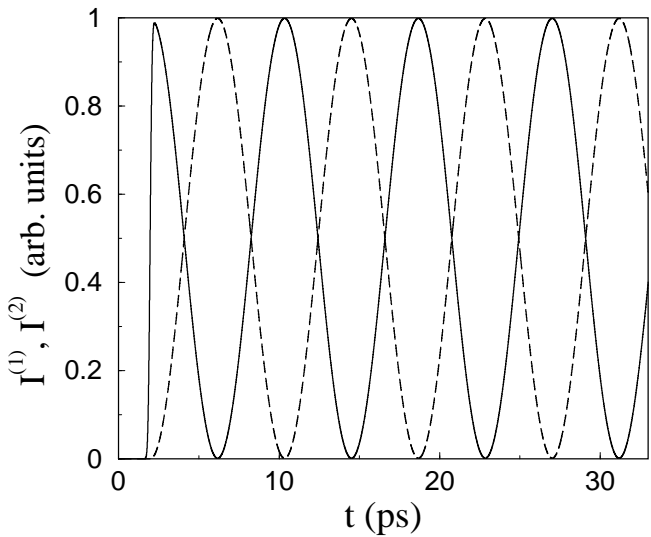}}
\caption{Time dependent oscillations of the integrated 
linear polarization intensities  
in the optically 
active medium. $I^{(1)}$ (solid line) and $I^{(2)}$ (dashed line),
see Eq. (\ref{ij}), 
are calculated for the pulse of Eqs. (\ref{ext}) and (\ref{ak}).
$x_0=-600$ $\mu$m, $\sigma_t=100$ fs, $\lambda_0=500$ nm, 
$\chi_{11}=0$, $\chi_{12}=0.0002$.}  
\label{4} 
\end{figure}

Figure \ref{4} shows the oscillation of the quantities 
\beq\label{ij}
I^{(j)}(t)\equiv \int_0^\infty dz\,|E^{(j)}(z,t)|^2,\;\;\; j=1,2,
\eeq
proportional to the total energy in each orthogonal linear polarization 
within the optically active medium 
versus time. For the parameters chosen, the pulse duration is much smaller
than the rotation period,   
so the oscillation is clearly visible. 
Whereas optical activity is standardly considered in coordinate space
and measured in stationary conditions at the end of a slab, 
here we adopt a different, time 
dependent view, closer to the quantum Rabi oscillation analog.   

\begin{figure}
{\includegraphics[width=3.3in]{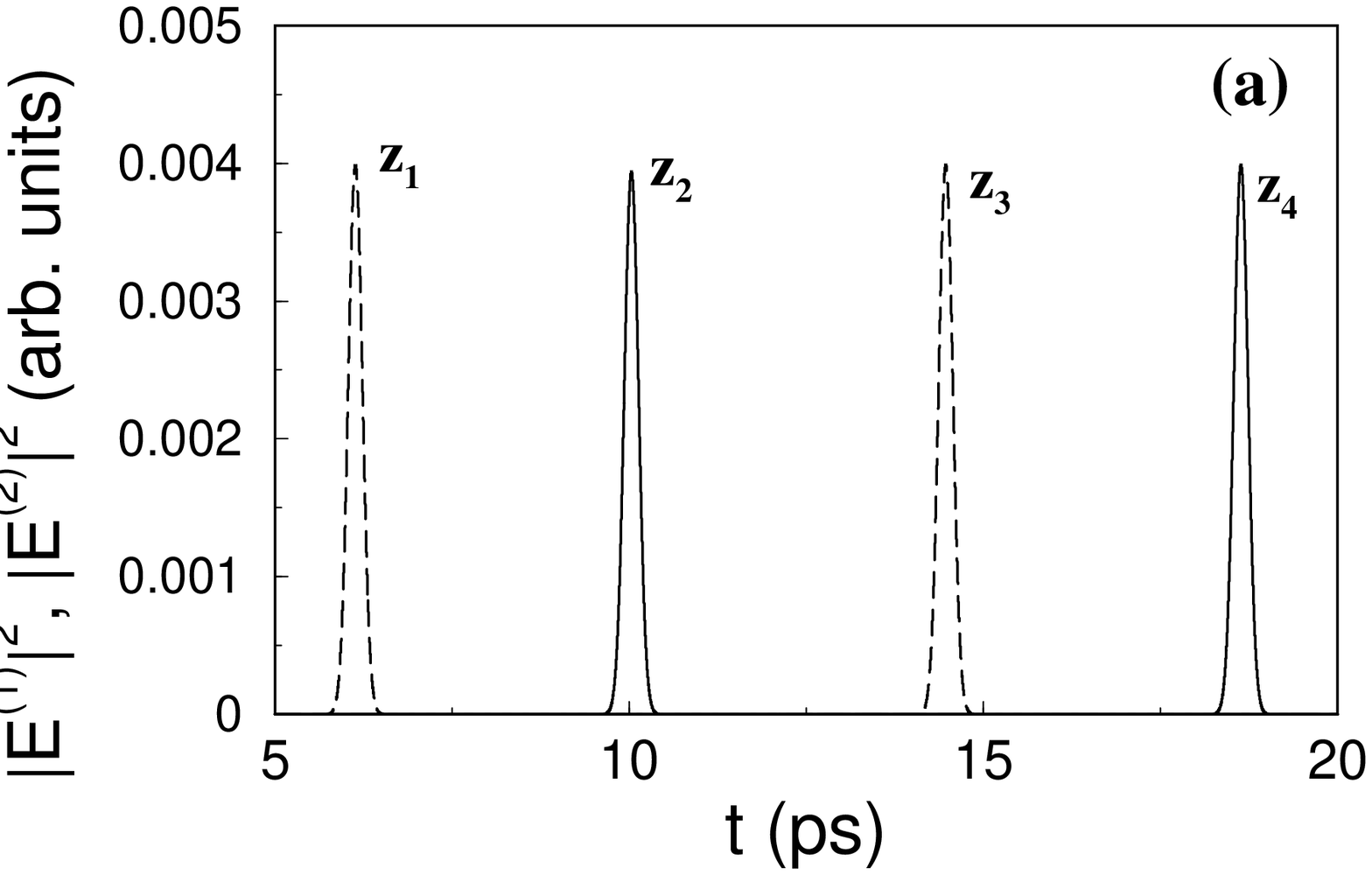}}
{\includegraphics[width=3.3in]{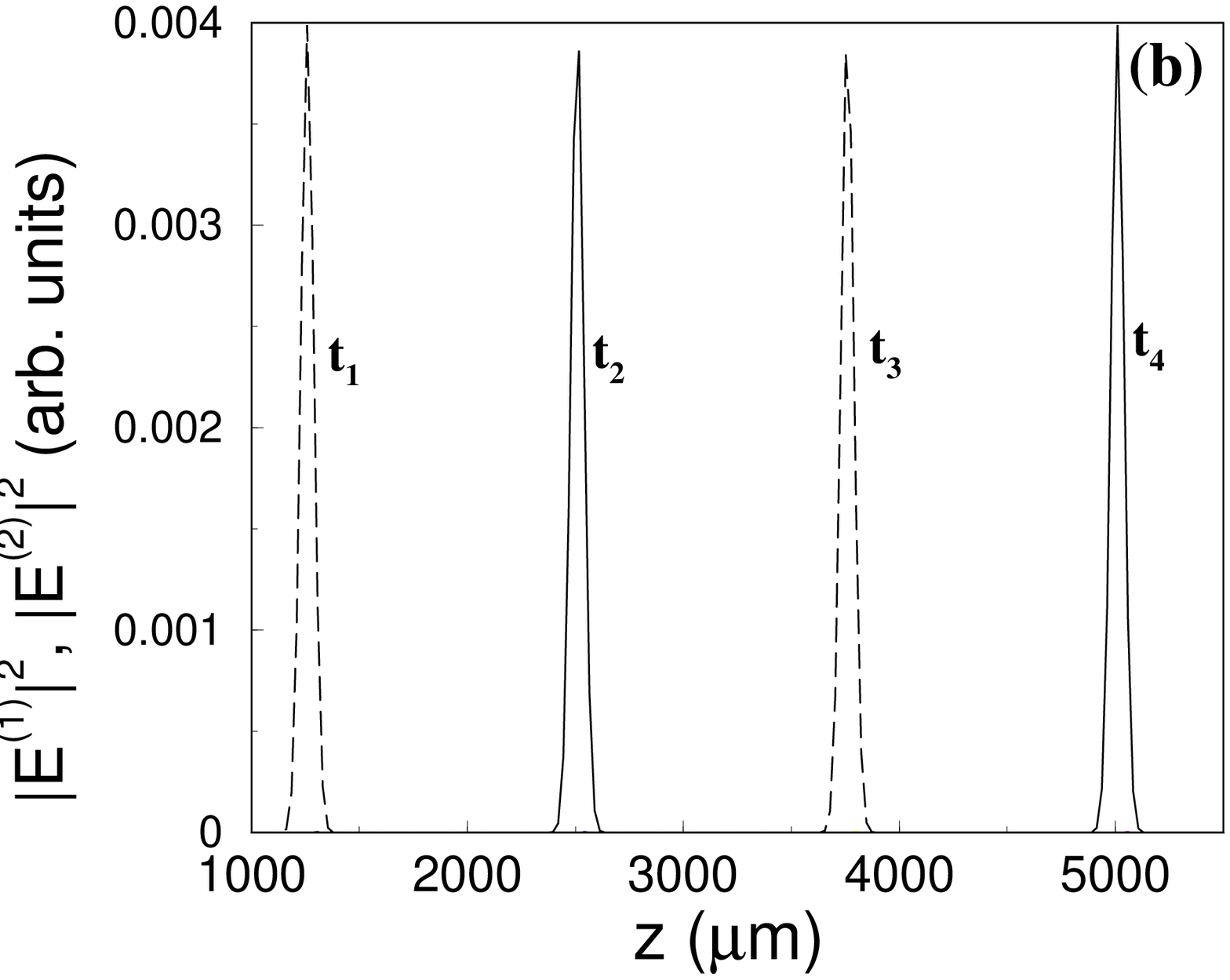}}
\caption{(a) The oscillations of the  
linear polarization components may also be seen 
at fixed positions separated by $L/2$:  
$z_1=1241$ $\mu$m, $z_2=2492$ $\mu$m, $z_3=3742$ $\mu$m,
$z_4=4993$ $\mu$m. The pulse arrives at these positions 
separated by time intervals $T/2$, totally dominated,
alternatively, by 
$|E^{(1)}|^2$ (solid line) or $|E^{(2)}|^2$ 
(dashed line). The parameters are the same as in Fig. \ref{4}. 
(b) Snapshots of the pulse taken at time intervals $T/2$: 
$t_1=6200$ fs, $t_2=10368$ fs, $t_3=14536$ fs, and $t_4=18704$ fs.
These times correspond to maxima in Fig. \ref{4}. 
As before, $|E^{(1)}|^2$ (solid line) or $|E^{(2)}|^2$ 
(dashed line) dominate the pulse alternatively 
at distances separated by $L/2$. Parameters as in Fig. \ref{4}.}   
\label{fotos} 
\end{figure}

Complementary perspectives of the time dependent oscillation
are provided in Fig. 
\ref{fotos} which represent measurements at fixed positions versus time,
or 
snapshots at fixed times of the intensities of the two 
orthogonal, linear polarization components
in arbitrary units.    

\begin{figure}
{\includegraphics[width=3.3in]{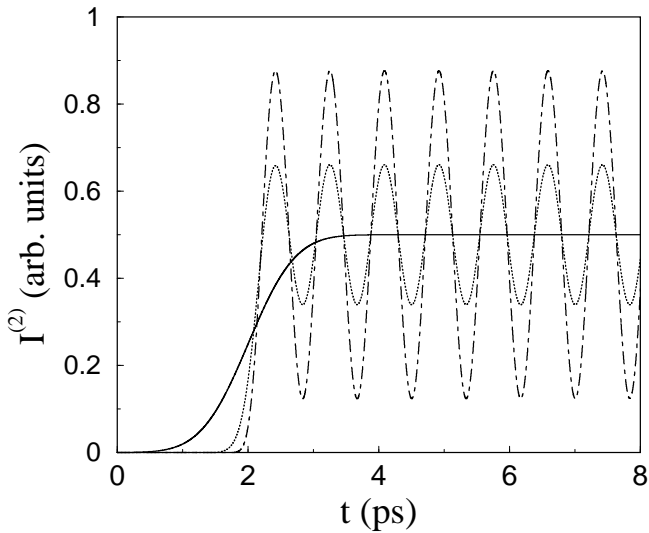}}
\caption{Oscillation suppression due to slow (adiabatic) entrance
in the optically active medium.  
$I^{(2)}(t)$ is represented for different pulse widths:   
$\sigma_t=100$ fs (dotted-dashed line); $\sigma_t=
200$ fs (dotted line); $\sigma_t=600$ fs (solid line). 
$\chi_{12}=0.002$. Other parameters as in Fig. 1.} 
\label{oscilacion8} 
\end{figure}
\begin{figure}
{\includegraphics[width=3.3in]{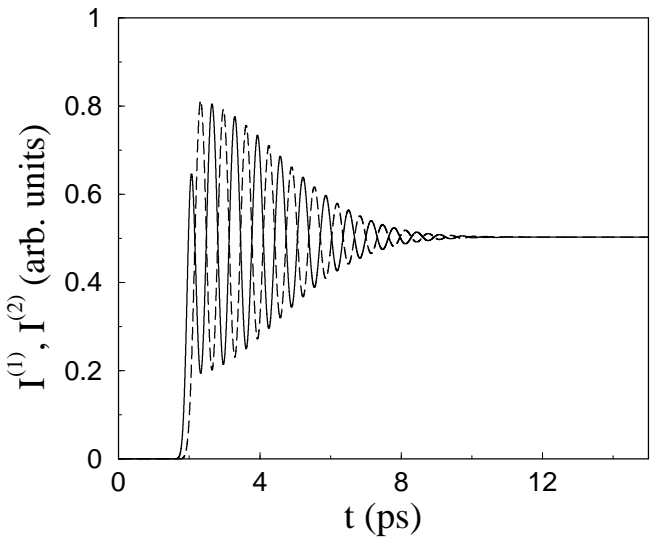}}
\caption{Oscillation suppresion of the integrated polarization 
intensities due to pulse splitting: 
$I^{(1)}$ (solid line) and $I^{(2)}$ (dashed line). 
$\chi_{12}=0.08$ and other parameters as in Fig. 1. 
A transition time that separates the two regimes (with or without rotation) 
may be estimated as
the time for the arrival of the pulse peak at the medium plus the 
splitting  time of Eq. (\ref{st}). This gives  
$4.5$ ps for the present parameters.}  
\label{51}
\end{figure} 
\begin{figure}
{\includegraphics[width=3.3in]{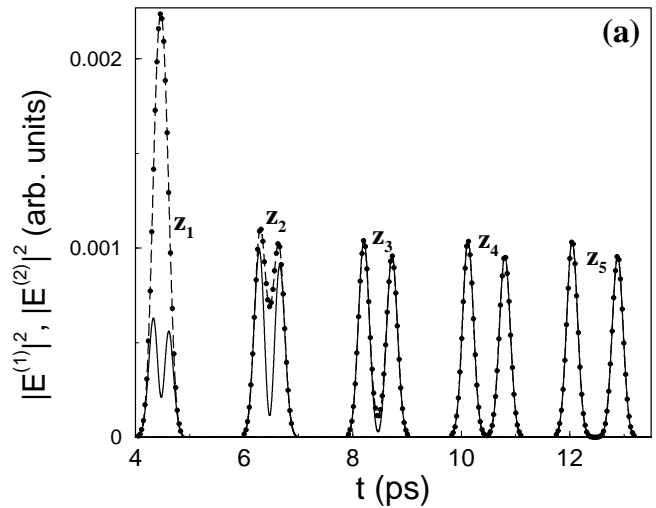}}
{\includegraphics[width=3.3in]{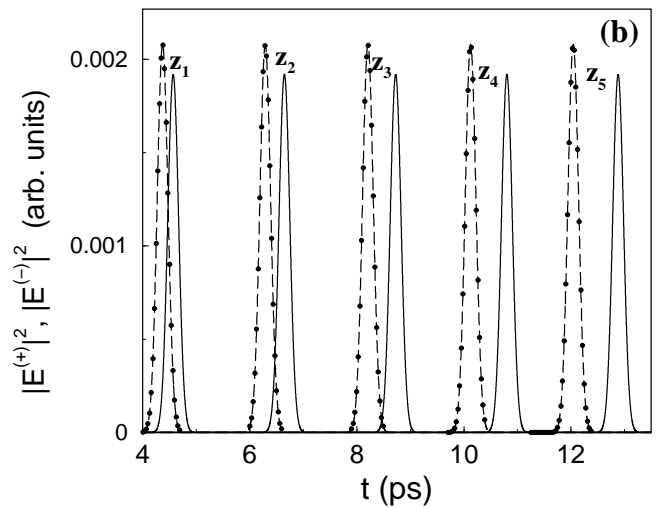}}
\caption{Modulus squared of 
polarization field components 
versus time at five different positions: $z_1=741$ $\mu$m, 
$z_2=1341$ $\mu$m, $z_3=1941$ $\mu$m, $z_4=2541$ $\mu$m, and 
$z_5=3141$ $\mu$m:  (a) vertical (solid line) and horizontal
(dashed line with dots) polarizations; (b)   
right (solid line) and left (dashed line with dots) 
circular polarizations.    
$\chi_{12}=0.08$. Other parameters as in Fig. 1.} 
\label{991}
\end{figure}
\begin{figure}
{\includegraphics[width=3.3in]{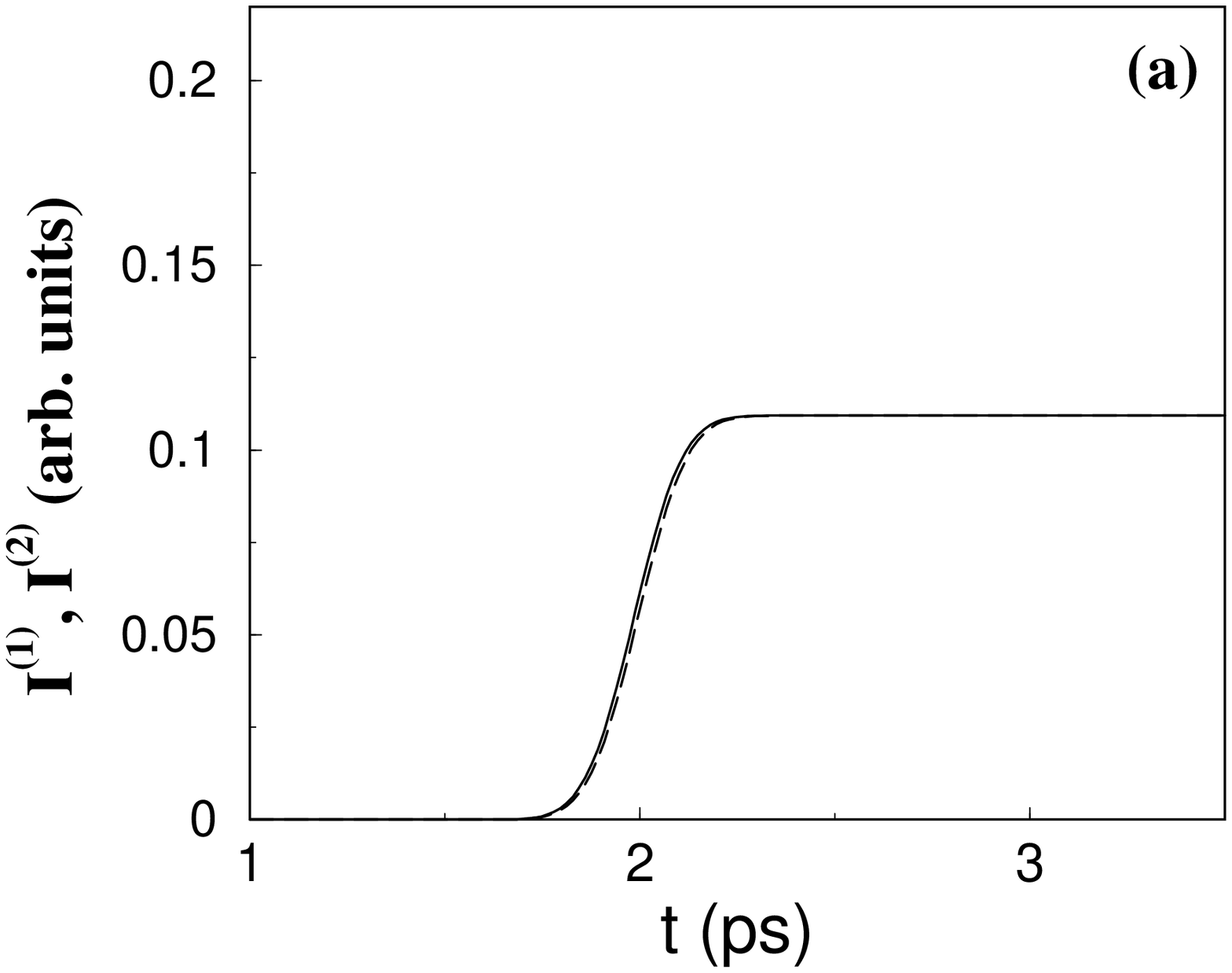}}
{\includegraphics[width=3.3in]{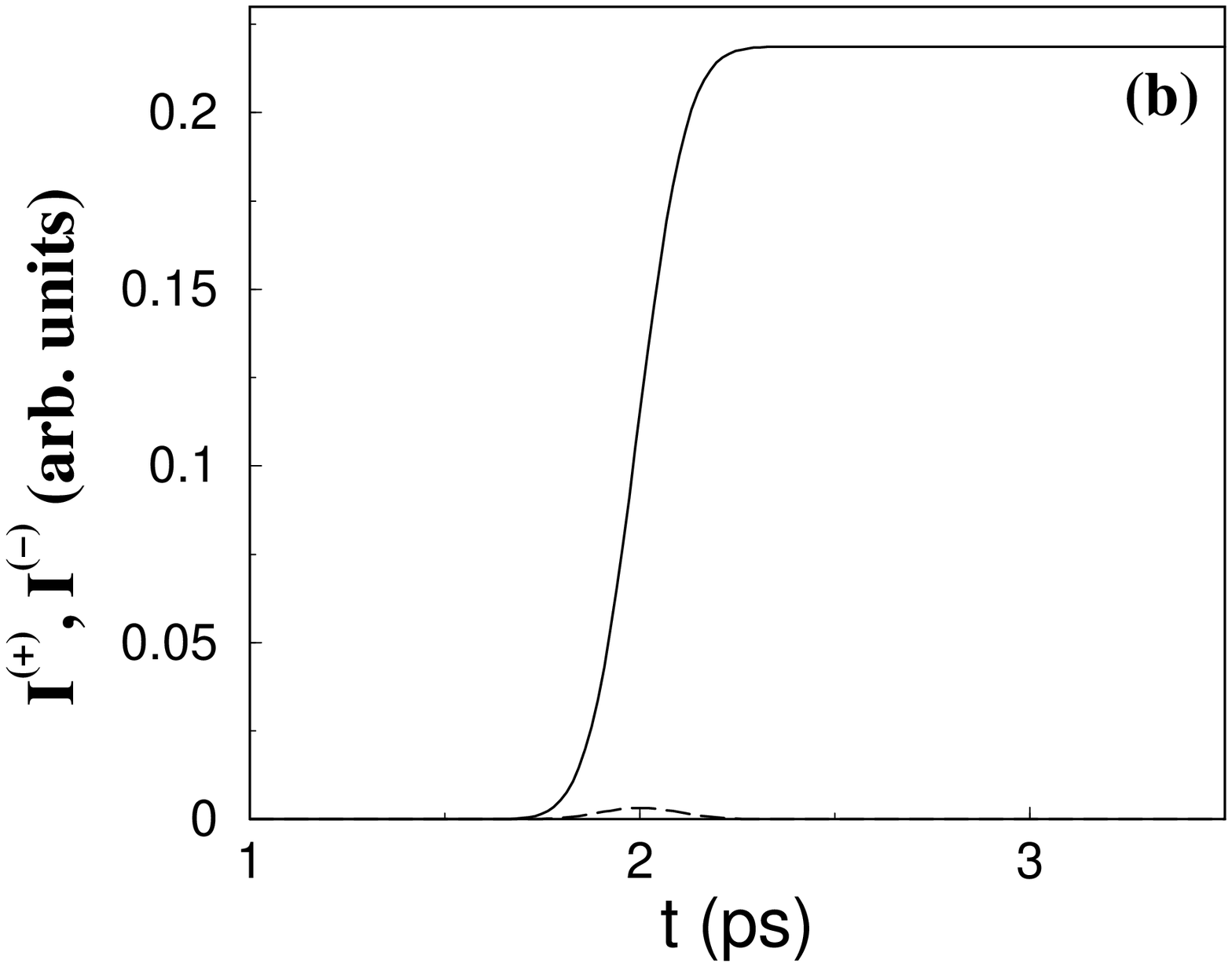}}
\caption{Oscillation suppresion of the integrated polarization 
intensities due to filtering. 
$\chi_{12}=1.2$ and other parameters as in Fig. 1.
In (a) we show the vertical (solid line) and horizontal (dashed line) 
polarization components
and in (b) the right (solid line) and left (dashed line) 
circular polarization components}  
\label{evan} 
\end{figure} 

If the pulse width is increased and becomes comparable or larger 
than the rotation wavelength, however, the oscillation is 
suppressed, as shown in Fig. \ref{oscilacion8}.
This is the optical analog of the adiabatic 
suppression of the Rabi oscillation due to a slow entrance of 
the atoms in the laser 
region. 

Other interesting phenomenon occurs if the observation time 
is large enough so that the pulse splits as a consequence 
of the two different group velocities. This is shown first in Fig. 
\ref{51}, 
where the oscillation eventually fades away because of the 
progressive lack of interference between the two circularly 
polarized components. A different view of this  
process is given in Fig. \ref{991}a, in which vertical and 
horizontal polarization components are represented versus time at
five different 
positions for the same pulse of Fig. \ref{51}. Notice that each linear 
polarization intensity develops two peaks,  
to be compared with the single peak pulses 
of  
Fig. \ref{fotos}. 

The gradual separation of the 
right and left circularly polarized components, 
which for $z>0$ are given by 
\beq
E^{(\pm)}\equiv\frac{1}{\pi^{1/2}}\int_0^\infty
 dk\, A(k) e^{-i\omega t} C_\pm e^{ik_\pm z},  
\eeq
is shown explicitly in 
Fig.
\ref{991}b. 
The pulse    
is finally formed by two distinct
peaks, each of them corresponding to a pure right or left 
circularly polarized subpulse. 

Finally, the rotation suppression by filtering, analogous to 
atomic state filtering, is illustrated in 
Fig. \ref{evan}. The left circular polarization becomes evanescent 
for the chosen susceptibility, $\chi_{11}=0$ and $\chi_{12}=1.2$, so,
after a transient, the pulse 
in the active medium is only composed by  
right handed circular polarization, 
see Fig. \ref{evan}b, 
where the total intensities  
\beq
I^{\pm}=\int_0^\infty dz\, |E^{(\pm)}|^2
\eeq
are represented versus time. 
The absence of left handed polarization 
after the transient peak  
precludes any oscillation of the 
linear polarization intensities $I^{(1,2)}$, as shown in Fig. 
\ref{evan}a.

\section{Magneto-optic effects\label{mo}} 

Instead of finding  materials  
with the susceptibility tensors 
necessary to observe the different
suppression effects, 
it is possible to manipulate $\bm{\chi}$ for an isotropic dielectric
by applying a static magnetic 
field ${\bf B}$. The susceptibility 
tensor matrix elements in terms of the resonance frequency
$\omega_0$, plasma frequency
$\omega_p$, and cyclotron frequencies $\omega_{cu}$, $u=x,y,z$, 
is given in the Appendix
for a simple Lorentz model. 
Let us first examine the ``Faraday configuration'', with the 
magnetic field along the $z$ direction. 
In that case the susceptibility takes the form given in 
Eq. (\ref{susfaraday}) and the solutions 
$k_\pm$ correspond to two orthogonal circular
polarizations.

\begin{figure}[h]
{\includegraphics[width=3.3in]{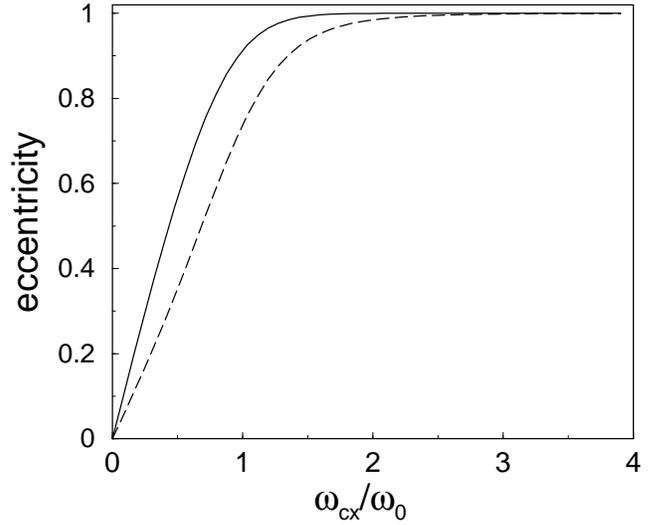}}
\caption{Eccentricity $e=\sqrt{\xi^2-\eta^2}/\xi$, where $\xi$ and $\eta$ are the 
semimajor and semiminor axes of the ellipses associated with the 
two orthogonal polarizations 
for $\omega/\omega_{0}=0.98$,
$0 \leq \omega_{cx}/\omega_{0} \leq 3.98$, $\omega_{cy}=0$, $\omega_{cz}/
\omega_{0}=2.65$. In this case $(E_y/E_x)_\pm$ is purely imaginary, 
so that 
the two semiaxes are 1 and $|(E_y/E_x)_\pm|$ \cite{BW}.}  
\label{exc2} 
\end{figure}

If instead the selected magnetic field direccion is
$x$ (Cotton-Mouton configuration), the form of the
susceptibility tensor becomes
\beq
{\bm{\chi}}=\left[
\begin{array}{ccc}
\chi_{11}&0&0
\\
0&\chi_{33}&\chi_{23}
\\
0&\chi_{23}^*&\chi_{33}
\end{array}
\right],\;\;\;{\rm{Re}}(\chi_{23})=0, 
\label{suscot}
\eeq
which leads to modes with linear polarizations in $x$ and $y$ directions. 
A field ${\bf B}$ in an arbitrary direction between the Faraday
and the Cotton-Mouton configurations produces, in general, two elliptic 
polarizations. Figure \ref{exc2} shows the eccentricities 
of the two polarizations starting from a Faraday 
configuration and ending up in a Cotton-Moutton configuration by 
increasing the $x$-component of the field. 

\subsection{Evanescence conditions}

Let us determine the conditions
which make one, and only one, of the propagation constants evanescent, 
to achieve polarization state filtering and selection. 
We shall work out in detail 
the Faraday configuration, but the Cotton-Mouton case 
or some other magnetic field orientation 
could be treated similarly. 
For the Faraday configuration,    
$k_\pm$ are either purely real or purely imaginary, 
and the parameter regions in which one wave becomes evanescent 
are delimited by the zeros 
of $k_\pm$, which  
correspond, respectively,   
to        
\beq\label{umc}
1+\chi_{11}=\mp |\chi_{12}|, 
\eeq
see Eq. (\ref{kmp}).  
Using the
expressions for $\chi_{11}$ and $\chi_{12}$ given 
in the Appendix, Eq. (\ref{umc}) becomes
\beqa\label{bra}
1\!+\!\frac{\omega_{p}^2}{[(\omega_{0}^2-\omega^2)^2-\omega^2 \omega_{cz}
^2]}\!=\!\pm {\bigg|}\!\frac{\omega_{p}^2 \omega \omega_{cz}}{[(\omega_{0}^2
-\omega^2)^2-\omega^2 \omega_{cz}^2]}\!{\bigg|},
\eeqa
with $\omega_{cz}=eB_z/m_e$. 
This leads to a
second order equations for the modulus
of $\omega_{cz}$,   
\beqa
-\!\omega^2\!{\it S}|\omega_{cz}|^2\!\pm\! \omega_{p}^2
\omega |\omega_{cz}|\!+\!
{\it S}\!(\!\omega_{0}^2\!-\!\omega^2\!)[\!(\!\omega_{0}^2\!
-\!\omega^2\!)\!+\!\omega_{p}^2]\!=\!0
\eeqa
where {\it S} the sign of $[(\omega_{0}^2-\omega^2)^2-\omega^2 \omega_{cz}
^2]$. 
The formal solutions are  
\beqa
\label{sol1}
\omega_{1cz}&=&\frac{-\omega_{p}^2-(\omega_{0}^2-\omega^2)}
{\omega}
\\
\label{sol2}
\omega_{2cz}&=&\frac{(\omega_{0}^2-\omega^2)}
{\omega}
\\
\label{sol3}
\omega_{3cz}&=&\frac{\omega_{p}^2+(\omega_{0}^2-\omega^2)}
{\omega}
\\
\label{sol4}
\omega_{4cz}&=&\frac{-(\omega_{0}^2-\omega^2)}
{\omega}
\eeqa
The parameter region in which filtering occurs ($k_-$ purely imaginary and 
$k_+>0$),   
is represented in Fig. 8 with a light shaded area (in the darker area both 
solutions are evanescent so there is total reflection).
It can be divided into four $\omega$-subregions 
I, II, III, IV, delimited by the curve crossing frequencies $\omega=\omega_0$,
$\omega=\omega_S\equiv
(\omega_0^2+\omega_p^2/2)^{1/2}$, and  $\omega=\omega_{0p}\equiv
(\omega_0^2+\omega_p^2)^{1/2}$. The lower and upper bounds
for the leftmost region,
I,  are the 
curves $\omega_{2cz}$ and $\omega_{3cz}$
(i.e., Eqs. (\ref{sol2}) and (\ref{sol3});
for region II,   
the curves $\omega_{4cz}$ and $\omega_{3cz}$; for region III, 
$\omega_{3cz}$ and $\omega_{4cz}$; 
and finally the bounds for region IV
are the curves $\omega_{1cz}$ and $\omega_{4cz}$. 
$\omega=\omega_S$ also marks a polarization mode change. 
In the light shaded areas, $(E_y/E_x)_\pm=\pm i$
if $\omega<\omega_S$, whereas 
$(E_y/E_x)_\pm=\mp i$ otherwise.      

An additional constraint is set by the maximum available intensity
of the magnetic field. It establishes a flat upper bound for $\omega_{cz}$,
and restricts the frequency range where filtering may be accomplished. 
The smallest fields could be used at or near 
the frequencies 
$\omega_0$ and $\omega_{0p}\equiv(\omega_0^2+\omega_p^2)^{1/2}$
where the boundary curves 
in Eqs. (\ref{sol1}-\ref{sol4}) touch the zero field axis. 
In practice the 
second one, $\omega_{0p}$,  
would be more useful since $\omega_p$ could be controlled in
some cases, e.g. for a gas. 
Moreover, in general the possible perturbation 
on the filtering effect by absorption 
can be made negligible   
if the condition $(\omega^2-\omega_0^2)>>\omega\gamma$ is satisfied. 
In $\omega_{0p}$ this assumes the form
\beq
{\omega_p^2}>>\gamma\omega_{0p},
\eeq
where $\gamma$ is the damping constant, see the Appendix. 
%
%
\begin{figure}
{\includegraphics[width=3.in]{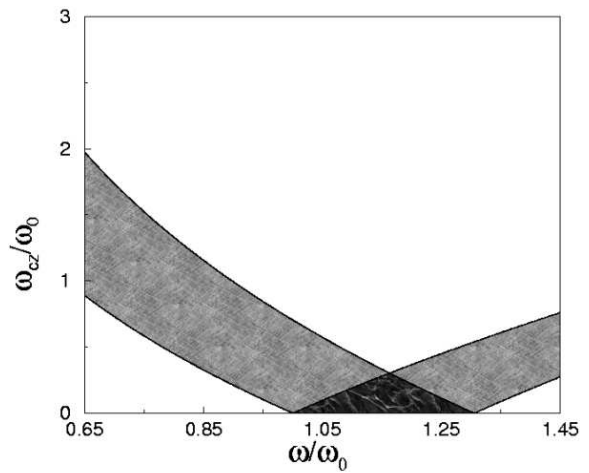}}
\caption{Parameter region in which $k_-$ is purely imaginary 
and $k_+>0$ (light shaded area) for  
$\omega_{p}/\omega_{0}=0.84$.
The critical points where curve 
crossings occur are,  
from left to right, $\omega$/$\omega_{0}$=$1$ 
(it separates regions I and II),
$\sqrt{1+(\omega_{p}^2/2\omega_{0}^2)}$ (between regions II and III),
and $\sqrt{1+(\omega_{p}^2/\omega_{0}^2)}$ (between regions III and IV).
}
\label{zonas}
\end{figure}
%
%

Two other important factors that must be taken into account for the
observability
and potential application  of the 
polarization filtering are the penetration length 
of the evanescent wave in the optically active medium, 
and the transmittance of the surviving mode.  

\begin{figure}[b]
{\includegraphics[width=3.5in]{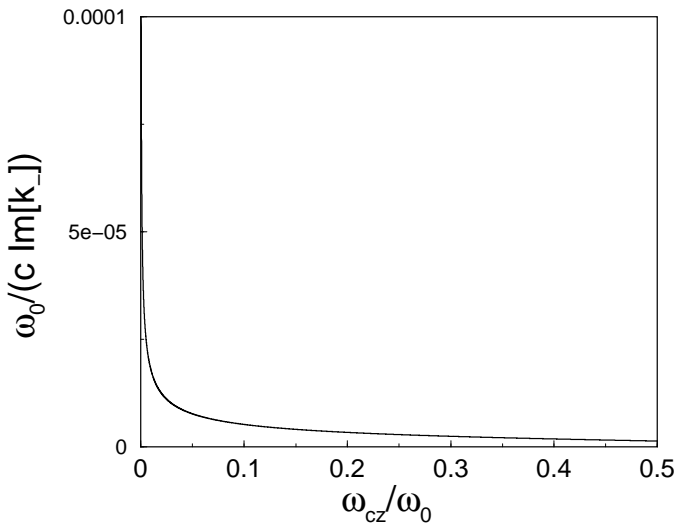}}
\caption{Dimensionless penetration length
of the evanescent wave at 
$\omega=\omega_{0p}$ for 
with $\omega_p/\omega_0=1$.
}
\label{long}
\end{figure}

The penetration length is
proportional to the inverse of $\rm{Im}(k_{-})$. 
In practice the medium is not 
semi-infinite, of course, but the filtering effect may be achieved for a 
finite medium as long as it extends beyond the penetration length.
This should not pose any practical problem according to the scales  
shown in Fig. {\ref{long}}, where $\omega_0/[c\,{\rm{Im}}(k_{-})]$ is shown 
versus $\omega_{cz}/\omega_0$ for $\omega=\omega_{0p}$.  

\begin{figure}
{\includegraphics[width=3.5in]{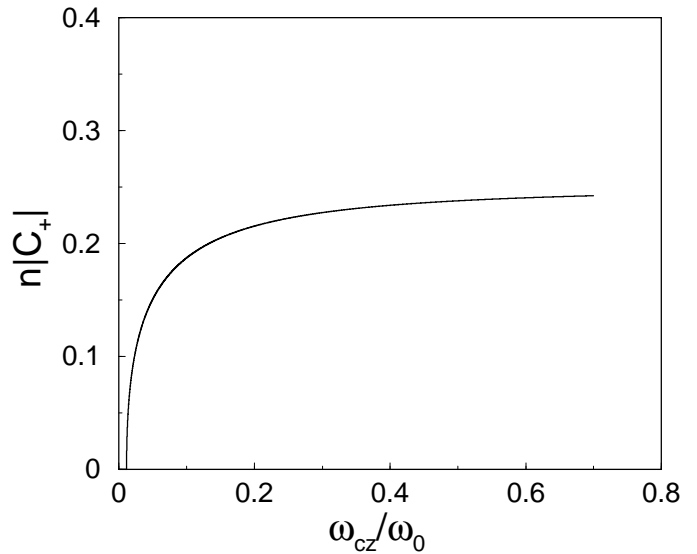}}
\caption{Transmittance of the surviving mode at $\omega=\omega_{0p}$
for $\omega_p/\omega_0=1$.}
\label{trans}
\end{figure}

The ``transmission probability'' $n_+|C_{+}|^2$ is the ratio between 
the energy flux of the surviving polarization and the inicident 
flux. 
It 
is showed in Fig. \ref{trans} for $\omega_0=\omega_p$.  
Up to a significant $25\%$ of transmission
may be obtained. 
Finally, real materials hold multiple resonances but our 
analysis can  
be easily generalized by summing over them with appropriate oscillator
strenghts factors \cite{Fowles}.

\section{Discussion}

In summary, we have shown that the atomic time-dependent Rabi oscillation 
and its suppression for moving atoms incident on a laser-illuminated 
region, is analogous to the time-dependent polarization rotation
and its suppression for    
light pulses entering into an optically 
active medium. The eigenmodes in the laser region depend on 
Rabi frequency and detuning and in the optically active medium on 
the susceptibility matrix elements. 
Since ultrashort femtosecond pulses can be tracked experimentally 
by interferometric photon scanning tunneling microscopy 
(PSTM) \cite{PSTM}, the time dependence of polarization 
rotation and its suppressions, 
by averaging, pulse splitting or filtering,  
can be tested experimentally. These effects may be of interest 
for the design of an all-optical
computer with information encoded, transfered 
or manipulated using the polarization state; 
similarly, their atomic counterparts  
may be relevant in metrology and provide a mechanism for controlled quantum
internal state preparation by projection or filtering, i.e., irrespective of the 
initial state, so that 
their experimental 
examination at a light-optics level is both feasible and 
worthwhile pursuing. 
The optical analog of atomic state-filtering 
provides a way to produce polarized light. Whereas the atomic 
state may be selected by 
playing with the laser detuning and intensity, the polarization 
may be selected 
by a magnetic field.

\begin{acknowledgments}
We are grateful to G. C. Hegerfeldt, D. Gu\'ery-Odelin,
A. Ruschhaupt, C. Salomon, and J. J. Gil   
for comments and encouragement, and to T. Pfau for his suggestion 
to examine an optical analog of Ref. \cite{NEMH03}.    
This work has been supported
by Ministerio de Ciencia y Tecnolog\'\i a (BFM2000-0816-C03-03
and HA2002-0002), and 
UPV-EHU (00039.310-13507/2001 and 15968/2004). 
\end{acknowledgments}

\appendix
\section{Susceptibility tensor elements \label{elem}}
%
The general expressions for the elements of the ${\bm{\chi}}$
tensor for a homogeneous dielectric in a magnetic field are 
obtained here from a simple, classical, Lorentz model. Each electron 
displacement from its equilibrium position ${\bf r}_e$ 
is assumed to satisfy 
\beq
m_e\frac{d^2 {\bf r}_e}{dt^2}=-e{\bf E}-e\frac{d{\bf r}_e}{dt}\times
 {\bf B}-K{\bf r}_e
-m_e\gamma\frac{d{\bf r}_e}{dt}
\eeq
where $K$ is an elastic force constant that keeps it bound, ${\bf B}$ is a 
static, external magnetic field, $m_e$ the mass of the electron 
and $\gamma$ a damping constant. 
(We neglect the small force due to the magnetic field of the optical wave).     
Assumming that the applied electric field and ${\bf r}_e$
vary harmonically as $e^{-i\omega t}$,
and using 
${\bf{P}}=-Ne{\bf r}_e={\bm{\chi}}\epsilon_0{\bf E}$ 
with $N$ the number of electrons
per unit volume, a lengthy but straighforward calculation 
gives, for $\gamma=0$,  
\beqa
\label{chis}
\chi_{11}&=&\omega_{p}^2 \frac{(\omega_{0}^2-\omega^2)^2 -\omega^2 
\omega_{cx}^2}{[(\omega_{0}^2-
\omega^2)^2 - \omega^2 \omega_{c}^2](\omega_{0}^2-\omega^2)}
\\
\chi_{12}&=&\omega_{p}^2 \frac{\omega[i\omega_{cz}(\omega_{0}^2-\omega^2)
 -\omega_{cx}\omega_{cy}\omega]}
{[(\omega_{0}^2-\omega^2)^2 -
 \omega^2 \omega_{c}^2](\omega_{0}^2-\omega^2)}
\nonumber\\
\chi_{13}&=&\omega_{p}^2 \frac{\omega[-i\omega_{cy}(\omega_{0}^2-\omega^2)
 -\omega_{cx}\omega_{cz}\omega]}
{[(\omega_{0}^2-\omega^2)^2 - 
\omega^2 \omega_{c}^2](\omega_{0}^2-\omega^2)}
\nonumber\\
\chi_{22}&=&\omega_{p}^2 \frac{(\omega_{0}^2-\omega^2)^2 -\omega^2 
\omega_{cy}^2}{[(\omega_{0}^2-
\omega^2)^2 - \omega^2 \omega_{c}^2](\omega_{0}^2-\omega^2)}
\nonumber\\
\chi_{23}&=&\omega_{p}^2 \frac{\omega[i\omega_{cx}(\omega_{0}^2-\omega^2)
-\omega_{cy}\omega_{cz}\omega]}
{[(\omega_{0}^2-\omega^2)^2 - \omega^2 \omega_{c}^2](\omega_{0}^2-\omega^2)}
\nonumber\\
\chi_{33}&=&\omega_{p}^2 \frac{(\omega_{0}^2-\omega^2)^2 -\omega^2 
\omega_{cz}^2}{[(\omega_{0}^2-
\omega^2)^2 - \omega^2 \omega_{c}^2](\omega_{0}^2-\omega^2)}
\nonumber
\eeqa
where 
\beqa
\omega_0&=&\sqrt{K/m_e},
\\
\omega_c&=&eB/m_e,\;\; \omega_{cu}=eB_u/m_e,\;\;u=x,y,z,
\\
\omega_p&=&\left(\frac{Ne^2}{m_e\epsilon_0}\right)^{1/2},
\eeqa
are resonance, cyclotron, and ``plasma'' frequencies, respectively, 
and $B=(B_x^2+B_y^2+B_z^2)^{1/2}$.  

In the hermitian case, i.e., for $\gamma=0$, $\chi_{ij}=\chi_{ji}^*$.     
If $\gamma\neq 0$ the elements in Eq. (\ref{chis}) would have
the same form except 
for the substitution 
\beq
\label{subs}
(\omega_{0}^2-\omega^2)\to(\omega_{0}^2-\omega^2-i\gamma\omega). 
\eeq
The other non-diagonal elements $\chi_{ij}$ can be obtained {\it formally}
by taking first the complex conjugate of the expressions of the transpose 
elements $\chi_{ji}$ in Eq. (\ref{chis}) and {\it then}
making the substitution 
of Eq. (\ref{subs}).

\end{document}